\documentclass[a4paper,twoside]{article}
\usepackage{AMS2LA,fancyhdr,CJK,xcolor,multicol,graphics,greekup,bm,cmap}
\usepackage{amssymb}\usepackage{gensymb,mathcomp}
\usepackage[pdfstartview=FitH, bookmarks=false, colorlinks=false,
    linkcolor=blue, urlcolor=blue, pdfborder=001, citecolor=blue,
    pdftitle={Changepdftitle}, pdfauthor={Changepdfauthor},
    pdfcreator=CHIN. PHYS. LETT., pdfproducer=CPL,
    pdfkeywords={changePACS}]{hyperref}
\newcommand{\explel}{{\large\sffamily~~~~~~~~~~~~~~~~~} \hfill}
\newcommand{\expler}{\hfill\fcolorbox{lightgray}{lightgray}{\textcolor{red}{{\large\fontfamily{phv}\textbf{New Submission}}}}}

\def\headrule{\kern 1mm \hrule width 17cm \kern -1mm}%
\def\footnoterule{\kern 1mm \hrule width 7cm \kern 2.2mm}%
\def\REF#1{\par\hangindent\parindent\indent\llap{#1\enspace}\ignorespaces}%

\newcommand{\ucite}[1]{$^{[#1]}$}
\renewcommand{\cite}[1]{\,[#1]}

\newcommand{\cplyear}{2022} \newcommand{\cplvol}{37}
\newcommand{\cplno}{x} \newcommand{\cplpagenumber}{xx{xxxx}}
 \newcommand{\cplpage}{\cplpagenumber-\thepage}

\usepackage{multirow}
\usepackage{graphicx}

\begin{document}
\begin{CJK}{UTF8}{gbsn}\vspace* {-6mm} \begin{center}


\large\bf{\boldmath{Measuring charge distribution of molecular cations by atomic Coulomb probe microscope}}
\footnotetext{\hspace*{-5.4mm}
	Supported by the National Basic Research Program of China (Grant No. 2019YFA0307700), the National Natural Science Foundation of China (Grants No. 12074143, 11974272, 11774281, and 12134005), and Science Challenge Project No. TZ2018005.

\noindent$^{*,\S}$Corresponding author. Email:  $^{*}$luosz@jlu.edu.cn; $^{\S}$dajund@jlu.edu.cn

\noindent\copyright\,{\cplyear}
\href{http://www.cps-net.org.cn}{Chinese Physical Society} and
\href{http://www.iop.org}{IOP Publishing Ltd}}
\\[6mm]

\normalsize \rm{}  
Xitao Yu(余西涛) $^{1}$, Xiaoqing Hu(胡晓青) $^{2}$, Jiaqi Zhou(周家琪)$^{3}$, Xinyu Zhang(张馨予) $^{1}$, Xinning Zhao(赵欣宁) $^{1}$, Shaokui Jia(贾少奎)$^{3}$, Xiaorui Xue(薛晓睿)$^{3}$,  Dianxiang Ren(任殿相) $^{1}$, Xiaokai Li(李孝开) $^{1}$, Yong Wu(吴勇) $^{2}$, Xueguang Ren(任雪光)$^{3}$, Sizuo Luo(罗嗣佐) $^{1,*}$, Dajun Ding(丁大军) $^{1,\S}$    
\\[2mm]\small\sl $^{1}$Institute of Atomic and Molecular Physics, Jilin University, Changchun 130012, China\\
$^{2}$Institute of Applied Physics and Computational Mathematics, Beijing 100088, China\\
$^{3}$School of Physics, Xi'an Jiaotong University, Xi'an 710049, China
\\[4mm]\normalsize\rm{}(Received  August 2022)
\end{center}
\end{CJK}
\vskip 1.5mm          
                      
\noindent{\narrower\small{}
	Imaging the charge distributions and structures of molecules and clusters will promote the understanding of the dynamics of the quantum system. Here, we report a method by using an Ar atom as a tip to probe the charge distributions of benzene (Bz) cations in gas phase. Remarkably, the measured charge distributions of Bz$^{+}$ ($\delta_{H} $ = 0.204, $\delta_{C} $ = -0.037) and Bz$ ^{2+}$ ($\delta_{H} $ = 0.248, $\delta_{C} $ = 0.0853) agree well with the calculated Mulliken distributions, and the structures of Bz$_{2}$ is reconstructed by using the measured charge distributions. The structures of two Bz$_{2}$ isomers (T-shaped and PD isomers) can be resolved from the measured inter-molecular potential $V(R)$ between two Bz ions, and the structures of Bz dimer agree well with the theoretical predictions. 
                      
\par}\vskip 3mm       
\normalsize\noindent{\narrower{PACS: 33.80.-h, 33.80.Gj, 34.20.Gj, 36.40.Mr}
{\rm\hspace*{13mm}DOI: 10.1088/0256-307X/39/11/113301}
\par}\vskip 6mm

Measurements of charge distributions and structures of molecules, clusters, and solids plays an essential role in physics, chemistry, and biology\ucite{1-4}. Charge distribution, or electron density distribution of a system reflects ability of atoms in a molecule to attract electrons. It is a basic assumption in the density functional theory model to deal with multi-electron problems\ucite{5,6}. Therefore, measurements of electron density distribution experimentally have always been dreams of physicists and chemists. Theoretically, charge distribution of a system can be given by the Mulliken population analysis\ucite{7}. Experimentally, electrostatic interaction between the tip and the sample can be successfully measured using Kelvin probe force microscopy to image the charge distribution in a single molecule attached to a thin insulating layer\ucite{8}.However, this requires the accurate work function of the sample and the tip. 

On the other hand, charge distribution of a system directly affects accuracy of structural imaging such as the Coulomb explosion (CE)\ucite{9-13}. As a well-known technique, Coulomb explosion imaging (CEI) has proven to be powerful in determining structures of weakly bonded clusters, even in specific quantum states\ucite{14-16}. Recently, the structures for small molecules, atomic clusters, and molecular dimers have been obtained\ucite{12-25}. However, only the distance between charge centers $ Rcc$ of fragments can be measured, suggesting that only the structures of atomic clusters can be determined accurately since the charge centers are the same as the mass centers\ucite{14-17}. In general, the charge center does not coincide with the mass center in a molecule, which makes it difficult to predict the structure of complex molecules and molecular clusters accurately. Therefore, the pre-determined charge distribution is of great significance for CEI to determine the atomic position of molecules accurately. In the atom-molecule cluster systems \ucite{13,21,22}, the atom can be directly used as a probe to determine the charge distribution of the adjacent molecular cations. Theoretically, for a molecule with known initial structure distribution, the space charge distribution of the molecule can be scanned in multi-dimension by atomic probes at different positions (atoms and molecules constitute dimer isomers) from their interaction energy, which can be obtained from the CEI method\ucite{17,24}. Therefore, questions remain as to whether it is possible to measure the charge distribution of molecular cations using the CE.

In this work, we establish an in-situ experimental method for measuring the charge distribution of molecular cation in the gas phase using an atom as the probing tip, i.e., an atomic Coulomb probe microscope (ACPM), as shown in Fig. 1. Using a benzene (Bz) molecule as a prototype, whose parameters describing the charge distribution are minimized, a single Ar atom can probe the charge distribution of the Bz cations, without any further multi-dimensional scanning. After measuring the charge distributions of Bz cations, the distributions have been applied to reconstruct the distances of mass center of $R_{BzBz}$ in the structure of (Bz)$_{2} $, which is a typical system to study the effects of $\pi-\pi$ and CH interactions between aromatic molecules. These interactions play important roles in stabilization of proteins, DNA, and solid materials containing aromatic groups\ucite{25-27}. The results show that the charge distribution and cluster structure can be determined from the measured kinetic-energy-release (KER) distribution of CE channels. This work marks a significant step towards measuring the charge distribution of molecular cation and imaging the structures of complex clusters.

	\vskip 4mm
	\centerline{\includegraphics[width=4.5in]{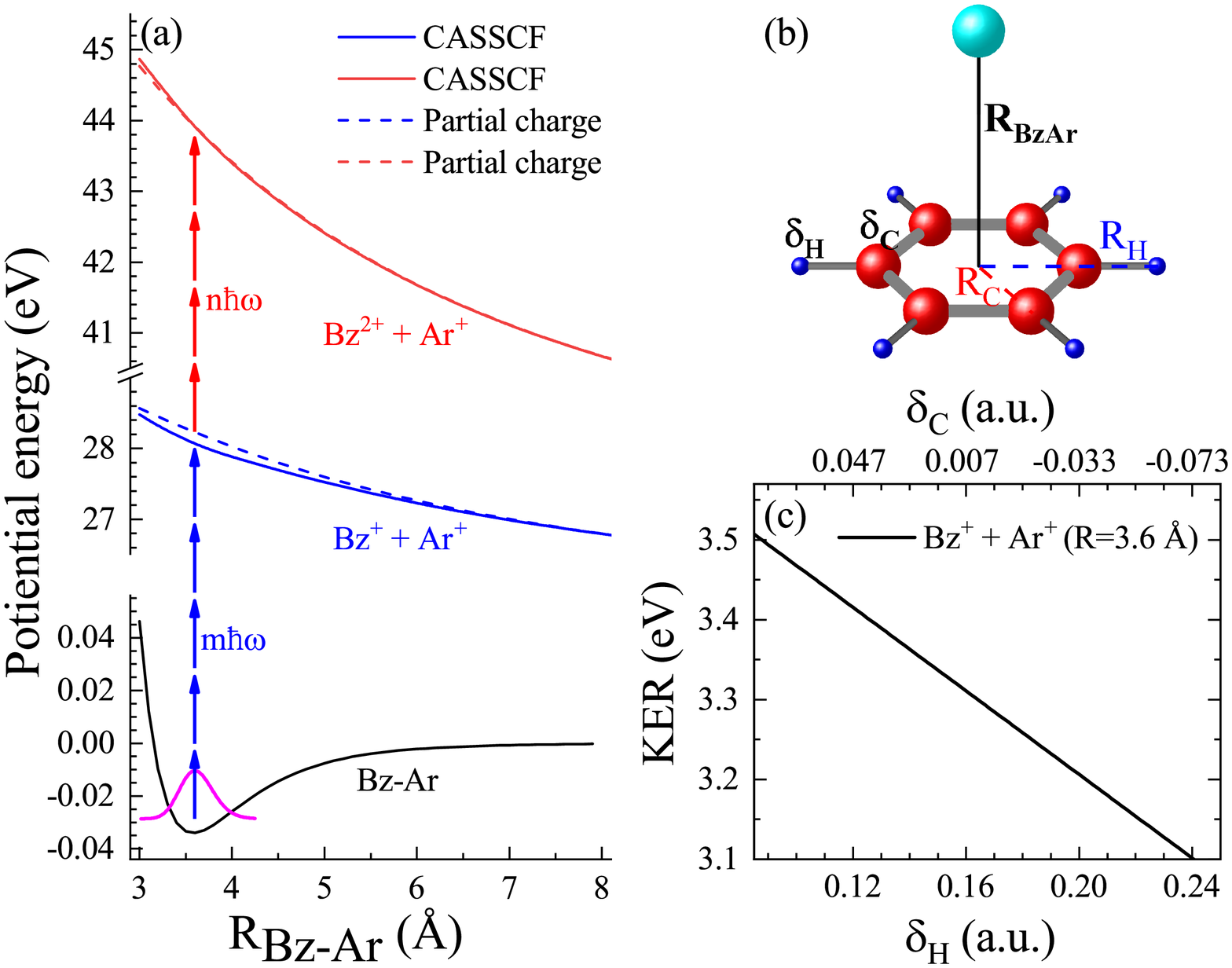}}
	\vskip 2mm
	\centerline{\footnotesize \begin{tabular}{p{4.5in}}\bf Fig.\,1. \rm
			The concept of charge distribution. (a) The corresponding PECs of the double and triple ionization of the BzAr calculated by CASSCF and partial charge model. (b) The schematic structure of BzAr with the coordinate and partial charge diagrams. (c) The  dependence of KER on $\delta_{H} $ and $\delta_{C} $ for the Bz$ ^{+} $ + Ar$ ^{+} $ channel with $R_{BzAr}$ = 3.6 \AA\ in partial charge model.	
	\end{tabular}}
	\vskip 0.5\baselineskip

 The ionization and CE pathways from femtosecond laser are shown in Fig 1(a). In the $Bz^{+}Ar^{+}$ and $Bz^{2+}Ar^{+}$ systems, the charge can be approximated as being concentrated on the atom as shown in Fig. 1(b), i.e., the charge is concentrated at the center-of-mass of the Ar atom, and the charge is distributed across the atoms in the Bz molecule as shown in Fig 1(b). Thus, the KER of these two channels accumulated from the Coulomb potential of $Bz^{+}Ar^{+}$ and $Bz^{2+}Ar^{+}$ in this partial charge model can be expressed as \ucite{28}	
	\begin{equation}
		KER = \frac{{6{\delta _H}}}{{\sqrt {R_{BzAr}^2 + R_H^2} }} + \frac{{6{\delta _C}}}{{\sqrt {R_{BzAr}^2 + R_C^2} }}
	\end{equation}
	where $R_{BzAr}$ is the distance between the mass centers of Bz and Ar, $R_H$ and $R_C$ are the distances from the H and C atoms to the mass centers of Bz which are set as the equilibrium positions of BzAr($R_C$=1.4004 \AA, $R_H$=2.4861 \AA), and $\delta_{H}$, $\delta_{C}$ are the partial charge distributed at H and C atoms with $6({\delta _{H}} + {\delta _{C}})$ equal to 1 and 2 for $Bz^+$ and $Bz^{2+}$. The corresponding potential energy curves (PECs) from quantum chemistry calculation \ucite{29} along with that obtained from the partial charge model in Fig. 1(a) show quantitative agreement, which indicates that the non-Coulomb and polarization effects between ions during the breaking of $Bz^{+}Ar^{+}$ and $Bz^{2+}Ar^{+}$ are weak compared to the CE energy, then, the KER of CE is well described by the partial charge model.  Thus, the precision measurements of the KER from the CE directly tag the charge distribution of the Bz cations as shown in Fig. 1(c). 
	
	The KER and correlations of the fragment ions from the CE channels are obtained by measuring the momenta of fragment ions in coincidence using COLTRIMS\ucite{30,31}. The experimental setup is depicted in Fig. S1(a) in the Supplemental Materials \ucite{32}, and more details have been given in our previous publications\ucite{11-13,33}. The BzAr and Bz$_2$ clusters are generated simultaneously in a supersonic expansion of benzene with argon as the carrier gas. The CE of different clusters is triggered by the laser field with the same parameters (800 nm, 35 fs, $~6\times10^{14}\ W/cm^2 $, ellipticity $\sim 0.95$), and by impacted with electrons (64 eV). Using the measured charge distribution of the Bz cations, we show that the isomers of the Bz dimers (T-shaped or parallel-displaced, PD) can be distinguished and their structures can be imaged.

	\vskip 4mm
	\centerline{\includegraphics[width=6.5in]{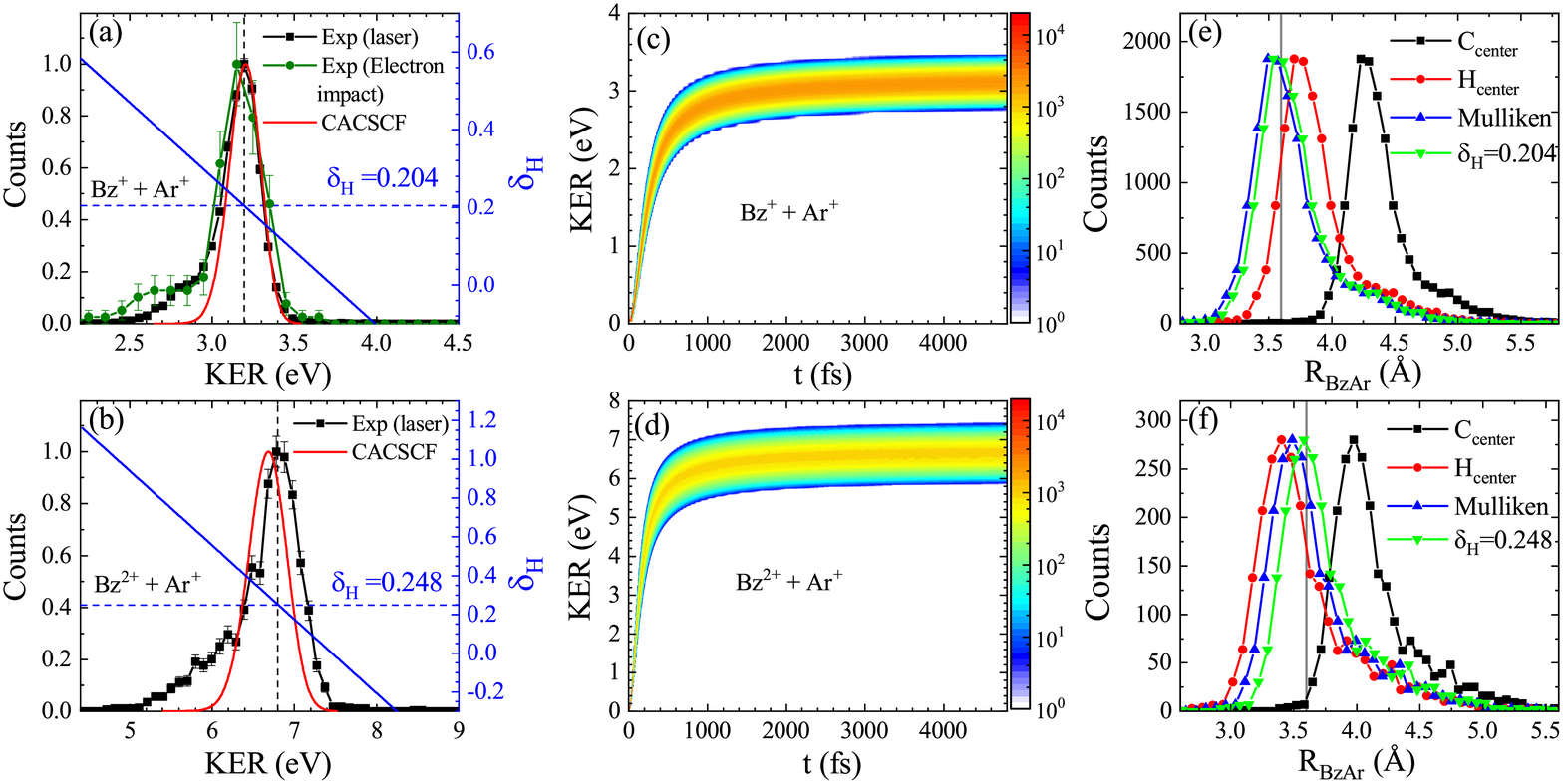}}
	\vskip 2mm
	\centerline{\footnotesize \begin{tabular}{p{6.5in}}\bf Fig.\,2. \rm
		The KERs of the channel (a) Bz$ ^{+} $ + Ar$ ^{+} $ and (b) Bz$ ^{2+} $ + Ar$ ^{+} $, with measurements from the laser-induced ionization plotted as black squares, measurement from electronic impact plotted as blue dots,  the simulated KER-dependent $\delta_{H} $ of the charge distribution plotted with red curves, assuming R = 3.6 \AA\ for BzAr. (c, d) The simulated time-dependent KER distribution of the breakup of these two channels. Panels (e) and (f) show the reconstructed distances R$ _{BzAr} $ from the two channels with four charge distributions; i.e., C-centered, H-centered, Mulliken distribution, and the measured distribution.			
	\end{tabular}}
	\vskip 0.5\baselineskip

	The measured KERs from the Bz$ ^{+} $ + Ar$ ^{+} $ and Bz$ ^{2+} $ + Ar$ ^{+} $ channels are plotted in Figs. 2(a) and (b). The most probable KERs of these two channels are 3.2 eV and 6.8 eV, and the corresponding distances Rcc between the charge centers are around 4.5 \AA\ and 4.2 \AA, both of which are larger than the predicted most probable distance between the mass centers, $R_{BzAr}$ $ \sim $ 3.6 \AA\ \ucite{34,35}, indicating that the center of mass does not coincide with the charge distribution. According to the partial charge model, the dependence of $\delta_{H}$ on KER for these two channels is given by the blue lines in Figs. 2(a) and 2(b). By comparing the most probable value of KER to the calculated curves, we obtain the partial charge of Bz$ ^{+} $ with $\delta_{H} =0.204$, $\delta_{C} = -0.037$, and of Bz$ ^{2+} $ with $\delta_{H} =0.248$, $\delta_{C} = 0.0853$.

   Charge distribution measurements can be affected by the structure deformation and charge migration during the ionization process, the polarization between ions, and the precision of the KER measurement. Since multiple ionization most likely occurs at the peak of the laser field, the time difference between the release of electrons in sequential ionization should be much smaller than the duration of a circularly polarized laser pulse\ucite{36}. To account for geometric deformation during a sequential multiple-ionization process, we have simulated the breakup process of Bz$ ^{+} $ + Ar$ ^{+} $ and Bz$ ^{2+} $ + Ar$ ^{+} $ by solving the relevant kinematic equations. In the Bz$ ^{+} $ + Ar$ ^{+} $ channel, $R_{BzAr}$ can safely be treated as a constant, since the amount of stretching is less than 0.0002 \AA\ when the propagation time is 20 fs\ucite{32}. Similarly, for the channel of Bz$ ^{2+} $ + Ar$ ^{+} $ at the same propagation time, the change in $R_{BzAr}$ is less than 0.04 \AA\ (seen in the Supplemental Materials). Thus, the influences of vibration during strong-field ionization can be neglected, and we can safely take the equilibrium distance with a fixed value of $R_{BzAr}$ = 3.6 \AA\ in the reconstructions. Furthermore, the measured KERs from strong-field ionization and electron impact agree with each other as shown in Fig 2(a), indicating that the influence of a strong field on structure deformation during CE can be neglected. Theoretically, the KERs of the two channels are accumulated from the calculated PECs from the CACSCF method with the initial distribution sampled in the ground state of BzAr, where the accumulations of KER with the evolution time for two channels are shown in Figs. 2(c) and 2(d) (the details can be seen in the Supplemental Materials). The most probable KERs are 3.2 and 6.7 eV for these two channels, which agree very well with the experiments. The influence of charge migration triggered by strong-field ionization is averaged by our CE measurement, since the electron migration of Bz$ ^{+} $ occurs at the fs to sub-fs time scale \ucite{37}, while the process of CE needs around 1 ps and 500 fs to accumulate the total KER for Bz$ ^{+} $ + Ar$ ^{+} $ and Bz$ ^{2+} $ + Ar$ ^{+} $ as given in Figs. 2(c) and 2(d). Thus, the averaged charge distributions of Bz cations are seen in this experiment.

	To check the measured charge distributions further, we made three assumptions at average Bz geometry: the charge is equally scattered across the six C atoms or the six H atoms, or the charge is given by the Mulliken distribution ($\delta_{H} $ = 0.212, $\delta_{C} $ = -0.046 for Bz$^{+}$ and $\delta_{H} $ = 0.295, $\delta_{C} $ = 0.038 for Bz$^{2+}$) obtained from quantum-chemistry calculations at the DFT/b3lyp/6-311+g(d,p) level\ucite{38}. The square of the vibrational wave function $ |\Psi(R_{BzAr})| ^{2} $ is then reconstructed from the previously known geometric characteristics of BzAr with different charge distributions, and the results are shown in Figs. 2(e) and 2(f). It is worthy to notice that the vibrational excitation of Bz$ ^{+} $ is ignored, which may contribute to the low energy side of KER and influence the $ |\Psi(R_{BzAr})| ^{2} $ distribution at large R$ _{BzAr}$ (seen in the Supplemental Materials). The most-probable distance R$ _{BzAr} $ from the reconstructions for the Bz$ ^{+} $ + Ar$ ^{+} $ channel equals 4.3, 3.8, and 3.5 \AA\ when the charge is located at the C atoms, the H atoms, or in the Mulliken distribution. In the other channel (Bz$ ^{2+} $ + Ar$ ^{+} $), the corresponding most-probable distances R$ _{BzAr} $ are 4.0, 3.4, and 3.5 \AA. The value of R$ _{BzAr} $ reconstructed from both channels using the Mulliken distribution agrees very well with the distance ($\sim$3.6 \AA) from previous predictions and observations as well as with the distance reconstructed by using the measured charge distribution. This indicates that the measured charge distributions agree well with the results predicated on quantum chemistry calculations. It is worthy to notice that the accuracy of our method depends on the agreement between the Coulomb potential and real potential, thus, our method will be working for systems with weak van der Waals and polarization interactions. Moreover, the neighbor ions induced charge redistribution by the polarization effect will be interesting to explore in the future studies.

	\vskip 4mm
	\centerline{\includegraphics[width=\linewidth]{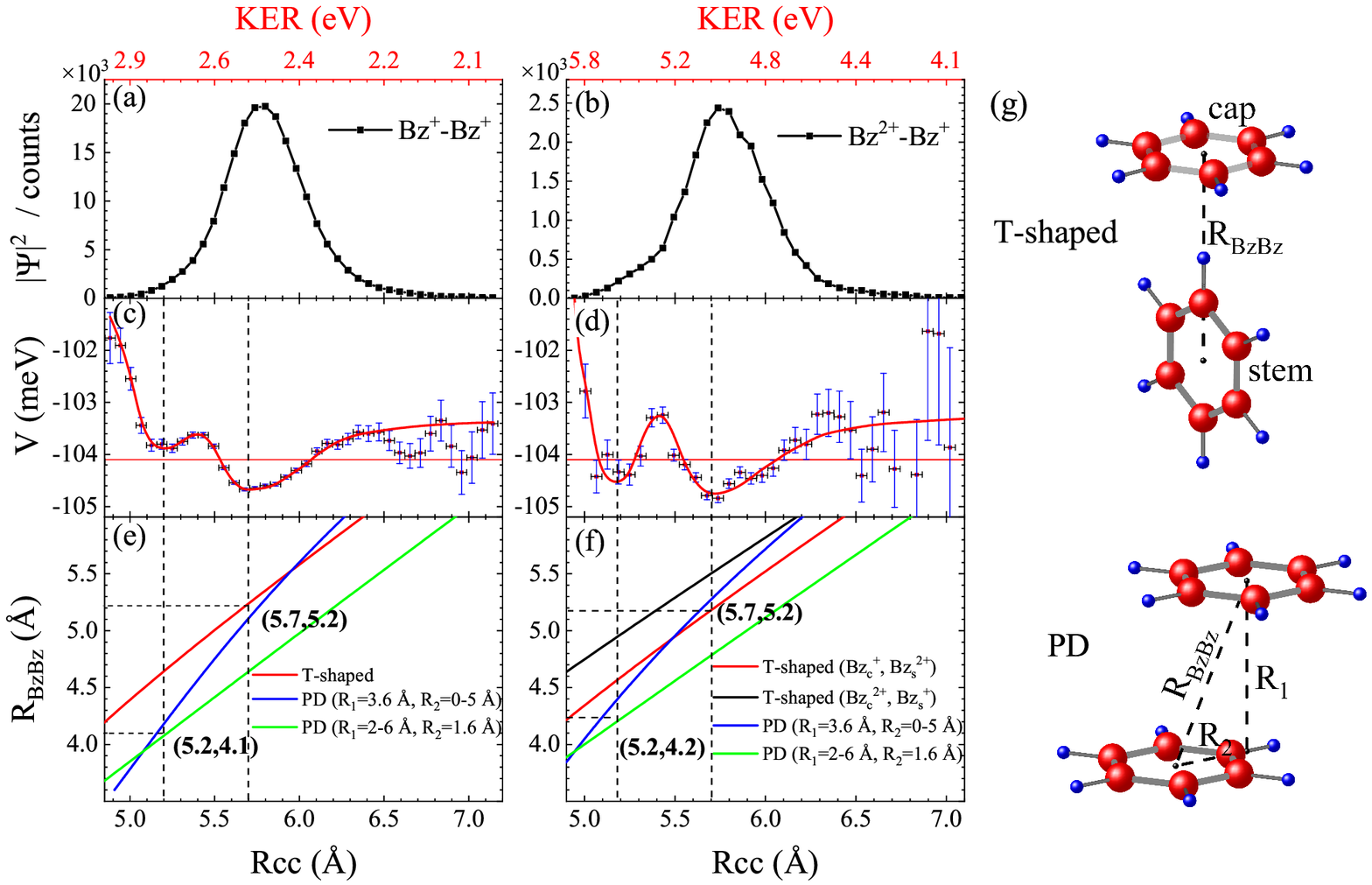}}
	\vskip 2mm
	\centerline{\footnotesize \begin{tabular}{p{6.8in}}\bf Fig.\,3. \rm
			Measured potential energy curve of (Bz)$ _{2} $, and the dependence of the  corresponding of R$ _{BzBz} $ on Rcc (the distance of average charge centers between two Bz) for (a,c,e) Bz$ ^{+} $ + Bz$ ^{+} $ and (b,d,f) Bz$ ^{2+} $ + Bz$ ^{+} $.  (a-b) The measured KER and $ |\Psi(R)|^{2} $ distributions and (c-d) The reconstructed intermolecular PECs from the two channels. The dependence of R$ _{BzBz} $ on Rcc from the two channels for the T-shaped and PD isomers are shown in panels (e) and (f). Two cases are modeled for the PD dimer; i.e., with R$ _{1} $ fixed at 3.6 \AA\ and R$ _{2} $ changing (blue lines) or with R$ _{1} $ changing and R$ _{2} $ fixed at 1.6 \AA\ (green lines). There are two possible charge configurations for Bz$ ^{+} $ + Bz$ ^{2+} $ channel; i.e., the stem Bz has double charge (Bz$ _{S^{2+}} $) and the cap Bz has single charge (Bz$ _{C^{+}} $) (red line) or the stem Bz is singly ionized (Bz$ _{S^{+}} $) and the cap Bz is doubly ionized (Bz$ _{C^{2+}} $) (black line).	(g) The structures of T-shaped and PD Bz dimers.			
	\end{tabular}}
	\vskip 0.5\baselineskip

	Based on the measured charge distribution of Bz cations and the CEI method, the structures of Bz clusters can be reconstructed from the KER of CE channels. For Bz dimers, the KER spectra of two CE channels, Bz$ ^{+} $ + Bz$ ^{+} $ and Bz$ ^{+} $ + Bz$ ^{2+} $, are shown in Figs. 3(a) and 3(b). The mean peak positions are 2.54 and 5.08 eV where the corresponding values of Rcc are around 5.7 \AA, which agrees well with the previously measured KER of the Bz$ ^{+} $ + Bz$ ^{+} $ with a peak value at 2.5 eV from electron impact experiment \ucite{39}. The measured profiles are determined by the square of the ground-state wave function, $|\Psi(R)|^2$. If the Bz dimer is considered approximately as a two-particle system with the given wave function $\Psi$(R), the interaction potential V(R) can be calculated from \ucite{17}
	\begin{equation}
		V\left(R\right) = \dfrac{\frac{\hbar^2}{2\mu}\dfrac{d^2\Psi\left(R\right)}{dR^2}}{\Psi\left(R\right)}+E,
		\label{eq2}
	\end{equation}
	where R is the effective distance between the two particles and E is the binding energy of the T-shaped Bz dimer (104 meV)\ucite{27,40}. The PECs of the two channels deduced from Eq. (2) are plotted in Figs. 3(c) and 3(d). Two minima appear at Rcc = 5.2 and 5.7 \AA\ for both the channels, which may result from the contribution of different isomers. The discrepancy between two minima in the PEC and one peak in the R distribution indicates that the PEC offers the possibility of distinguishing the structures of isomers with low populations since the PEC depends on the second derivative of the wave function and its value is related to the curvature rather than the amplitude of $|\Psi(R)|^2$.
	
	The measured KER spectra show that the Coulomb energy from the channel Bz$ ^{2+} $ + Bz$ ^{+} $ is twice the value of Bz$ ^{+} $ + Bz$ ^{+} $, which indicates that  the redistribution induced by the neighbor ions only has a weak effect and no influences for different charge states are observed. Thus, we can connect the intermolecular distance R$ _{BzBz} $ and Rcc for the two isomers of the Bz dimer (the T-shaped and PD dimers) by using the measured charge distributions. The results are shown in Figs. 3(e) and 3(f). In the Bz$ ^{+} $ + Bz$ ^{+} $ channel, the dependence of R$ _{BzBz} $ on Rcc for the T-shaped dimer is shown with the red line in Fig 3(e), and the most probable value of R$ _{BzBz} $ equals 5.2 \AA\ when Rcc is 5.7 \AA. In the Bz$ ^{+} $ + Bz$ ^{2+} $ channel, there are two possible charge configurations for the T-shaped dimer, as shown in Fig. 3(g); i.e., the stem Bz is double charged (Bz$ _{s} $$^{2+}$) and the cap Bz is single charged (Bz$ _{c} $$^{+}$), or else the stem Bz has a single charge (Bz$ _{s} $$^{+}$) and the cap Bz has a double charge (Bz$ _{c} $$^{2+}$). The most probable value of R$ _{BzBz} $ from the reconstruction equals 5.2 \AA\ in the first case for the Bz$ ^{+} $ + Bz$ ^{2+} $ channel, which agrees with the value of 5.2 \AA\ from the Bz$ ^{+} $ + Bz$ ^{+} $ channel (the red line in Figs. 3(e) and 3(f)). However, the value of R$ _{BzBz} $ is 5.5 \AA\ for the second charge configuration of the Bz$ ^{+} $ + Bz$ ^{2+} $ channel, as shown by the black line in Fig. 3(f). Since there is an agreement for the first charge configuration, it indicates the stem Bz is doubly ionized and only one electron is ionized from the cap Bz. 
	
	Furthermore, the measured value R$_{BzBz}$ around 5.2 \AA\ correlated with the T-shaped dimer agrees well with the theories\ucite{27,40,41}, and the corresponding KER at 2.53 eV agrees well with the previous \textit{ab initio} molecular dynamics simulations of the KER for the T-shaped isomer at 2.6 eV \ucite{39}. For the PD dimer, we modeled two cases with R$ _{1} $ fixed at 3.6 \AA\ and R$ _{2} $ changing or with R$ _{1} $ changing and R$ _{2} $ fixed at 1.6 \AA. In both cases, we reach a value of R$ _{BzBz} $ around 4.1 \AA\ when Rcc equals 5.2 \AA. Thus, the dimer structure corresponding to the minimum at Rcc = 5.2 \AA\ can be assigned to the PD structure, where the reconstructed R$ _{BzBz} $ are around 4.1 and 4.2 \AA\ from two channels and these values agree with the theoretical predictions \ucite{27,40,41}. The corresponding KER of this minimum is 2.8 eV, which agrees well with the \textit{ab initio} molecular dynamics simulations\ucite{39}. This is the first experimental observation of the structure of the PD isomer.

		\begin{table*}[!htb]
		\begin{center}	
			\parbox{5.7in}{
				\caption{
					The reconstructed and predicated distance R between the mass centers, assuming that the charge is distributed in the mass centers (M$ _{c} $), C centers (C$ _{center} $), H centers (H$ _{center} $), the Mulliken distribution, and the measured distribution.
				}
			}
	
			\setlength{\tabcolsep}{1.7mm}{
				\begin{tabular}{cccccccc}
					\hline\hline
					\multicolumn{2}{c}{\multirow{2}*{channel}}  	& \multicolumn{6}{c}{R (\AA)} \\  \cline{3-8}
					& 				 		& M$_{c}$ & C$_{center}$ & H$_{center}$ & Mulliken & Measured  & Theory \\  \hline
					\multicolumn{2}{c}{Bz$^{+} $ + Ar$^{+}$ }  		 & 4.5 	& 4.3	& 3.8 & 3.5 &   		& \multirow{2}*{3.53/3.58\textsuperscript{\ucite{34,35}} } \\
					\multicolumn{2}{c}{Bz$^{2+} $ + Ar$^{+}$ } 		& 4.2 	& 4.0& 3.4 & 3.5  &   		\\   \hline
					\multirow{2}*{T-shaped} & Bz$^{+}$ + Bz$^{+}$  	& 5.7 	& 5.6 	& 5.3 & 5.2 & 5.2  	& \multirow{2}*{5.0/5.1\textsuperscript{\ucite{27,40,41}} }  \\
					& Bz$^{2+}$ + Bz$^{+}$ 	& 5.7	& 5.6 	& 5.3 & 5.2 & 5.2  	\\	\hline
					\multirow{2}*{PD} 	& Bz$^{+}$ + Bz$^{+}$  		& 5.2 	& 5.0 	& 4.3 & 4.0 & 4.1 	& \multirow{2}*{3.94/4.02\textsuperscript{\ucite{27,40,41}} } \\
					& Bz$^{2+}$ + Bz$^{+}$ 		& 5.2	& 5.0 	& 4.3 & 4.1 & 4.2 	\\   \hline
				
					\hline\hline
					
			\end{tabular}}
		\end{center}
	\end{table*}

	The measurements of the density of the vibrational wave function $|\Psi(R)|^{2}$ can provide well-defined initial conditions for studying intermolecular processes such as bimolecular reaction and excimer formation, as well as determine the intermolecular PECs, as shown in Figs. 3(c) and 3(d). These measurements prove that T-shaped and PD dimers coexist in the molecular beam and demonstrate the interconversion between two isomers is important\ucite{42}. Furthermore, the reconstruction of V(R) demonstrates the ability to distinguish isomers even when their populations are too low to be discriminated directly from the measured KERs, which provides a tool for tracking the evolution between isomers. The measured R for BzAr, and Bz dimer from different channels, together with theoretical results, are summarized in Table I. The comparisons demonstrate that the CE can be used to determine the structures of a complex molecular cluster, and this method can also be performed by high-energy electron and ion collisions and triggered by high-energy photons from synchrotron radiation and free-electron lasers \ucite{39,43,44}.

	In conclusion, our study provides a new \textit{in-situ} ACPM method for measuring the charge distributions of molecular cations experimentally. The charge distributions of Bz cation ($\delta_{H} $ = 0.204, $\delta_{C} $ = -0.037) and dication ($\delta_{H} $ = 0.248, $\delta_{C} $ = 0.0853) are measured by using an Ar atom as the probing tip, and the results agree with quantum-chemistry calculations. As an important application, the geometric configurations of Bz isomers are resolved and imaged, \textit{i.e.}, T-shaped (R$ _{BzBz}\sim $ 5.2 \AA) and PD (R$ _{BzBz}\sim $ 4.1 \AA, 4.2 \AA) Bz isomers. Furthermore, the measured potential V(R) of the Bz dimer extracted from the curvature of the wave function demonstrates the ability to distinguish isomers of the Bz dimer. Our method provides possibility to track the ultrafast charge transfer between atoms and molecules during the molecular breaking and intermolecular Coulombic decay processes \ucite{45,46}, which will promote our understanding on the dynamics charge arrangement in the small quantum systems. These results pave the way to imaging the charge migration and nuclear dynamics of clusters with $fs$ temporal and sub-\AA\ spatial resolution.

\section*{\Large\bf References}

\vspace*{-0.8\baselineskip}\frenchspacing

\hskip 7pt {\footnotesize

	\REF{[1]}	 
		Neutze R, Wouts R, van der Spoel D, Weckert E and Hajdu J 2000 \textit{Nature} \textbf{406} 752
	\REF{[2]}	 
		Ihee H, Lorenc M, Kim T K, Kong Q Y, Cammarata M, Lee J H, Bratos S, and M. Wulff 2005 \textit{Science} \textbf{309} 1223
	\REF{[3]}	
		Gaffney K J and Chapman H N 2007 \textit{Science} \textbf{316} 1444
	\REF{[4]}	
		 Gross L, Mohn F, Moll N, Liljeroth P, and Gerhard Meyer 2009 \textit{Science} \textbf{325} 1110
    \REF{[5]} 
        Parr R G 1983 \textit{Annu. Rev. Phys. Chem.} \textbf{34} 631--56
    \REF{[6]} 
        Geerlings P, De Proft F and Langenaeker W 2003 \textit{Chem. Rev.} \textbf{103} 1793
    \REF{[7]} 
        Mulliken R S 1955 \textit{J. Chem. Phys.} \textbf{23} 1833 
	\REF{[8]}	 
        Mohn F, Gross L, Moll N and Meyer G 2012 \textit{Nat. Nanotechnol.} \textbf{7} 227
	\REF{[9]}
		 Pitzer M, Kunitski M, Johnson A S, Jahnke T, Sann H, Sturm F, Schmidt L Ph H, Schmidt-B\"ocking H, D\"orner R, Stohner J, Kiedrowski J, Reggelin M, Marquardt S, Schie{\ss}er A, Berger R and Sch\"offler M S 2013 \textit{Science} \textbf{341} 1096
	\REF{[10]}
		 Herwig P, Zawatzky K, Grieser M, Heber O, Jordon-Thaden B, Krantz C, Novotny O, Repnow R, Schurig V, Schwalm D, Vager Z, Wolf A, Trapp O and Kreckel H 2013 \textit{Science} \textbf{342} 1084
	\REF{[11]}
		 Zhao X, Yu X, Xu X, Yin Z, Yu J, Li X, Ma P, Zhang D, Wang C, Luo S and Ding D 2020 \textit{Phys. Rev. A} \textbf{101} 013416
	\REF{[12]}
        Yu X, Zhao X, Wang Z, Yang Y, Zhang X, Ma P, Li X, Wang C, Xu X, Wang C, Zhang D, Luo S and Ding D 2021 \textit{Phys. Rev. A} \textbf{104} 053104
    \REF{[13]}
        Yu X, Liu Y, Deng K, Zhang X, Ma P, Li X, Wang C, Cui Z, Luo S and Ding D 2022 \textit{Phys. Rev. A} \textbf{105} 063105
	\REF{[14]}
		 Kunitski M, Zeller S, Voigtsberger J, Kalinin A, Schmidt L Ph H, Sch\"offler M, Czasch A, Sch\"ollkopf W, Grisenti R E, Jahnke T, Blume D and D\"orner R 2015 \textit{Science} \textbf{348} 551
	\REF{[15]}
		 Voigtsberger J, Zeller S, Becht J, Neumann N, Sturm F, Kim H-K, Waitz M, Trinter F, Kunitski M, Kalinin A, Wu J, Sch\"ollkopf W, Bressanini D, Czasch A, Williams J B, Ullmann-Pfleger K, Schmidt L P H, Sch\"offler M S, Grisenti R E, Jahnke T and D\"orner R 2014 \textit{Nat. Commun.} \textbf{5} 5765
	\REF{[16]}
		 Li X, Yu X, Ma P, Zhao X, Wang C, Luo S and Ding D 2022  \textit{Chinese Phys. B} \textbf{31} 103304
	\REF{[17]}
		 Zeller S, Kunitski M, Voigtsberger J, Waitz M, Trinter F, Eckart S, Kalinin A, Czasch A, Schmidt L Ph H, Weber T, Sch\"offler M, Jahnke T and D\"orner R 2018 \textit{Phys. Rev. Lett.} \textbf{121} 083002	 	 
	\REF{[18]}
		 Pickering J D, Shepperson B, H\"ubschmann B A K, Thorning F and Stapelfeldt H 2018 \textit{Phys. Rev. Lett.} \textbf{120} 113202
    \REF{[19]}
        Yu X, Zhang X, Hu X, Zhao X, Ren D, Li X, Ma P, Wang C, Wu Y, Luo S and Ding D 2022 \textit{Phys. Rev. Lett.} \textbf{129} 023001
	\REF{[20]}
		 Ulrich B, Vredenborg A, Malakzadeh A, Schmidt L Ph H, Havermeier T, Meckel M, Cole K, Smolarski M, Chang Z, Jahnke T and D\"orner R 2011 \textit{J. Phys. Chem. A} \textbf{115} 6936
	\REF{[21]}
		 Wu C, Wu C, Song D, Su H, Xie X, Li M, Deng Y, Liu Y and Gong Q 2014 \textit{J. Chem. Phys.} \textbf{140} 141101
	\REF{[22]}
		 Wu J, Kunitski M, Schmidt L P H, Jahnke T and D\"orner R 2012 \textit{J. Chem. Phys.} \textbf{137} 104308
	\REF{[23]}
		 Song P, Wang X, Meng C, Dong W, Li Y, Lv Z, Zhang D, Zhao Z and Yuan J 2019 \textit{Phys. Rev. A} \textbf{99} 053427
	\REF{[24]}
		 Khan A, Jahnke T, Zeller S, Trinter F, Sch\"offler M, Schmidt L Ph H, D\"orner R and Kunitski M 2020 \textit{J. Phys. Chem. Lett. }\textbf{11} 2457
	\REF{[25]}
		 Burley S and Petsko G 1985 \textit{Science }\textbf{229} 23
	\REF{[26]}
		 Hunter C A and Sanders J K M 1990 \textit{J. Am. Chem. Soc.} \textbf{112} 5525
	\REF{[27]}
		 Sinnokrot M O, Valeev E F and Sherrill C D 2002 \textit{J. Am. Chem. Soc.} \textbf{124} 10887
	\REF{[28]}
		 Schouder C A, Chatterley A S, Madsen L B, Jensen F and Stapelfeldt H 2020 \textit{Phys. Rev. A} \textbf{102} 063125
    \REF{[29]}
		Werner H J, Knowles P J, Knizia G, Manby F R, Sch\"otz M, Celani P, Korona T, Lindh R, Mitrushenkov A, Rauhut G et al., MOLPRO, a package of ab initio programs, version 2010.1, See http://www.molpro.net.
	\REF{[30]}
		 D\"orner R, Mergel V, Jagutzki O, Spielberger L, Ullrich J, Moshammer R and Schmidt-B\"ocking H 2000\textit{ Phys. Rep.} \textbf{330} 95
	\REF{[31]}
		 Ullrich J, Moshammer R, Dorn A, D\"orner R, Schmidt L P H and Schmidt-B\"ocking H 2003\textit{ Rep. Prog. Phys.} \textbf{66} 1463
	\REF{[32]}
		See Supplemental Material for more details about the experimental setup, the simulations, and additional data from experiments and the quantum chemistry calculations.
	\REF{[33]}
		 Luo S, Liu J, Li X, Zhang D, Yu X, Ren D, Li M, Yang Y, Wang Z, Ma P, Wang C, Zhao J, Zhao Z and Ding D 2021 \textit{Phys. Rev. Lett.} \textbf{126} 103202
	\REF{[34]}
		 Capelo S B, Fern\'andez B, Koch H and Felker P M 2009 \textit{J. Phys. Chem. A} \textbf{113} 5212
	\REF{[35]}
		 Esteki K, Barclay A J, McKellar A R W and Moazzen-Ahmadi N 2018 \textit{Chem. Phys. Lett.} \textbf{713} 65
	\REF{[36]}
		 Pfeiffer A N, Cirelli C, Smolarski M, D\"orner R and Keller U 2011 \textit{Nat. Phys.} \textbf{7} 428
	\REF{[37]}
		 Despr\'{e} V, Marciniak A, Loriot V, Galbraith M C E, Rouz\'{e}e A, Vrakking M J J, L\'{e}pine F and Kuleff A I 2015 \textit{J. Phys. Chem. Lett.} \textbf{6} 426
	\REF{[38]}
		 Frisch M J Gaussian 09 Revis. B01 2010 Gaussian Inc. Wallingford CT 2009	
    \REF{[39]}
        Ren X, Zhou J, Wang E, Yang T, Xu Z, Sisourat N, Pfeifer T and Dorn A 2022 \textit{Nat. Chem.} \textbf{14} 232-8
	\REF{[40]}
		 Sinnokrot M O and Sherrill C D 2006 \textit{J. Phys. Chem. A} \textbf{110} 10656
	\REF{[41]}
		 Jha P C, Rinkevicius Z, \AA gren H, Seal P and Chakrabarti S 2008 \textit{Phys. Chem. Chem. Phys.} \textbf{10} 2715
	\REF{[42]}
		 \v Rez\'a\v c J and Hobza P 2008 \textit{J. Chem. Theory Comput.} \textbf{4} 1835
    \REF{[43]} 
        Ren X, Wang E,  Skitnevskaya A D,  Trofimov A B,  Gokhberg K,  Dorn A. 2018 \textit{Nat. Phys.} \textbf{14} 1062. 
	\REF{[44]}
	    Rudenko A, Inhester L, Hanasaki K, Li X, Robatjazi S J, Erk B, Boll R, Toyota K, Hao Y, Vendrell O, Bomme C, Savelyev E, Rudek B, Foucar L, Southworth S H, Lehmann C S, Kraessig B, Marchenko T, Simon M, Ueda K, Ferguson K R, Bucher M, Gorkhover T, Carron S, Alonso-Mori R, Koglin J E, Correa J, Williams G J, Boutet S, Young L, Bostedt C, Son S K, Santra R and Rolles D 2017 \textit{Nature} \textbf{546} 129	  
	 \REF{[45]}
	     Erk B, Boll R, Trippel S, Anielski D, Foucar L, Rudek B, Epp SW, Coffee R, Carron S, Schorb S and Ferguson KR 2014 \textit{Science} \textbf{345} 288.	  
	 \REF{[46]}
	  	Zhou J, Yu X, Luo S, Xue X, Jia S, Zhang X, Zhao Y, Hao X, He L, Wang C, Ding D, and Ren X 2022 \textit{Nat. Communications} \textbf{13} 5335
	  
}

\end{document}